\documentclass[5p]{elsarticle}
\usepackage[hyphens]{url}
\usepackage[hidelinks]{hyperref}
\usepackage{fancyvrb}
\usepackage{xspace}
\usepackage{enumitem}
\usepackage{paralist}
\usepackage{siunitx}
\newcommand{\be}{BE\xspace}
\newcommand{\sbuf}{\emph{sbuf}\xspace}
\newcommand{\sbufs}{\emph{sbuf}s\xspace}
\journal{a conference}
\newcommand{\tabref}[1]{Table~\ref{#1}}

\newcommand{\secref}[1]{\S\ref{#1}}

\begin{document}
\begin{frontmatter}

\title{Sharpening Your Tools: Updating {\tt bulk\_extractor} for the 2020s.}

\author{Simson Garfinkel}
\address{Digital Corpora Project}
\author{Jonathan Stewart}
\address{Aon Cyber Solutions}


\begin{abstract}
Bulk\_extractor is a high-performance digital forensics tool written
in C++. Between 2018 and 2022 we updated the program from C++98 to
C++17, performed a complete code refactoring, and adopted a unit test
framework. The new version typically runs with 75\% more throughput
than the previous version, which we
attribute to improved multithreading. We provide lessons and
recommendations for other digital forensics tool maintainers.
\end{abstract}


\end{frontmatter}


\section{Introduction}
Digital forensics (DF) is a fast moving field with a huge subject area. A digital
investigator must be able to analyze ``any data that might be found on
any device anywhere on the planet.''\cite{garfinkel:american-scientist-digital-forensics} As such,
DF tools must be continually updated to address new
file formats, new encoding schemes, and new ways that the subjects of
an investigation (the ``targets'') use their computers. At the same time,
tools need to retain the ability to analyze legacy data formats---all of
them, in fact.

Many DF tools run on
stock operating systems (Linux, macOS and Windows), adding another layer
of complexity: operating systems are also continually evolving.
Analysts who do not update their systems risk having those systems
compromised by malware, which negatively impacts analyst productivity and can be used in court to discredit an analysis. This is true even for analyst workstations
that are ``air gapped'' and not connected to the Internet, since
malware in evidence can exploit bugs in forensic software\cite{celebrite-zeroday}. At the same
time, updating the operating system potentially causes risk for the
proper execution of the digital forensic tools: although
new versions of operating systems attempt to provide
compatibility for software that ran on previous versions,
compatibility layers are not perfect.

Surprisingly, open source software distributed in source code form faces a greater challenge when
the underlying operating system is upgraded. This is because software
compatibility layers typically emphasize compatibility for software
that is distributed as  binary executables. Software that is
compiled from source, in contrast, must cope with upgrades to
compilers, libraries, and file locations. Old compilers required for
old source code distributions may not run on new operating
systems. Legacy software may use older libraries that are
incompatible with newer runtimes. Thus, after the compilers and
libraries are upgraded, the older open source software may no longer
compile.
Colloquially this is sometimes called \emph{dependency
hell} and \emph{bit rot}. One way around this problem is to run the old software inside a
virtual machine---but older virtual machines typically won't be
protected against modern malware threats.

In theory, one advantage of open source software is that the end-user has the source
code and is therefore able to update the application (or pay for a
programmer to update the application). In practice, many users of
digital forensic tools lack the expertise, financial resources, and
 time to update the collection of open source tools that they rely
upon to do their jobs.
\subsection{Contribution}

This article presents our experience updating the digital forensics tool \texttt{bulk\_extractor} (BE)\cite{garfinkel:bulk-extractor}
a decade after its initial release. Forensic tool developers can benefit from the detailed discussion of how embracing features in the C++17 standard and modern software engineering practices can improve the correctness, reliability, and throughput of forensic software. Businesses and funding agencies can use this experience to help justify the substantial cost of updating and even rewriting digital forensic tools that appear to be working properly. Students will benefit from reading this article and then consulting the BE source code, which can be found on GitHub.

\subsection{Outline}
This concludes the introduction. In \secref{background} we present a detailed description of BE, including the tool's history and
its use in digital forensics research and education. We  discuss
why it is difficult to ascertain the extent that an open source digital
forensics tool is used operationally.\footnote{Although \be version 1.5.3
is the most recent version of the program that was widely in use at
the start of this project, v1.5.3 would
no longer compile on modern systems. Therefore we had to separately
update v1.5.3, producing versions 1.6 and 1.6.1, in addition to our rewrite that
produced v2.0.}

In \secref{improvement} we present how we planned the update of the
tool. Included in this section is a discussion of the improvements in
the C++ and Python programming languages over the past decade that
might inspire others to upgrade their tools. We also discuss the
design and inclusion of a test suite, including both unit tests and
black-box tests.

In \secref{validation} we present the results of our
effort to produce \be version 2.0 (BE2), including refactoring the
code base and implementing a significant number of unit tests that
were used to validate the correctness, reliability and
throughput improvements of the tool.

Finally, in \secref{recommendations} we generalize our experience in
updating BE and extract lessons that may be useful for
the developers of other forensic applications.

\section{Background}\label{background}

Although there are many models for digital evidence examination, the
most common include a five step process of policy and capability
development; evidence assessment; evidence acquisition; evidence
examination; documentation and reporting\cite{199408}. BE is designed
to assist in the evidence examination stage.

Multiple strategies are employed by evidence examination tools. There are file-extraction tools that attempt to extract individual files from disk images or reassemble files from network streams using metadata; there are file carving tools, which attempt to recognize files within bulk data such as disk image and product files based solely on content recognition; and there are file analysis tools that understand file formats and attempt to extract information, such as text and Microsoft Office file metadata.

BE does not fit neatly into any of these categories. Instead, it was
designed to be a so-called ``find evidence button.'' It is similar to a
file carving tool, in that it attempts to recognize known formats in bulk data and use those
data in further processing. But in addition to recognizing files such as JPEG images, BE will recognize smaller ``features'' such as email addresses, URLs and credit card numbers, as such information has been proven to be valuable in investigations.  BE will also examine every
input block to see if that block contains FAT32 or NTFS directory
entries and, if any are found, report the decoded metadata. Overall, it handles
dozens of data formats, all at the same time. The program then
constructs normalized Unicode histograms of important strings like email addresses and
Internet search queries, drawing from utf-8, utf-16be and utf-16le encoded
binary.
Experience has shown that this ``kitchen-sink'' approach---throwing
every tool at every byte---finds data that other tools miss, and
these data can be important in investigations.
And while such analysis is computationally expensive, it is embarrassingly parallel. As a result, BE routinely utilizes all of the
cores of a multi-core workstation.

Another distinguishing aspect of BE is that it performs recursive reanalysis of data
blocks. That is, BE checks every byte to see if it is the start of a
stream that can be decompressed, decoded, or otherwise unwrapped. If
so, the resulting bytes are then recursively reanalyzed using  the
various \be scanners. Thus, BE's JPEG carver finds not just ordinary
JPEGs, but JPEGs that are in gzip-compressed data, JPEG's that are in
BASE64 MIME attachments, and JPEGs that were once in Windows memory
but have since been compressed and written to the Windows swap file.

The combination of decoding data recursively and recognizing interesting
data without regard to file system structure makes \be a powerful tool
that complements traditional forensics tools. Because of its simplified
data model and parallelism, \be can typically process a disk image several
times faster than traditional tools, and can therefore be used for triage. BE also supports a random-sampling mode, making it possible, for example, to scan a 1TB disk image in 5 minutes and determine with 99\% probability if the disk contains a specific media file from an archive.\cite{8433959} At the same time, \be sometimes unearths artifacts that other
tools can't, making it useful for demanding investigations.

Because BE ignores file boundaries, the modules that it uses to
recognize content, called \emph{scanners}, are typically more complex
than the format decoders (sometimes called \emph{dissectors}) in other
forensic programs. Of course each scanner checks the input to every field before
using it for memory references. But BE scanners also check for
end-of-memory conditions, since a scanner may be operating on a
fragment of a decompressed memory block. And since BE processes
memory in parallel, each block in a different thread, all scanners
must be thread-safe. Some of the program's most important scanners are large
lexical analyzers written in GNU Flex\cite{Paxson:1988:FFL} that scan bulk data for email
addresses, phone numbers, MAC addresses, IP addresses, URLs, and other kinds of
formatted text strings (sometimes called \emph{selectors}\cite{lawfare:selectors}). The approach of using
GNU flex for this purpose was first used by  SBook\cite{sbook} to recognize email addresses, phone
numbers, and other formatted information in free-text address book
entries; the BE scanners are based on the original SBook analysis engine, meaning that some of the code in BE is now 30 years old.

\subsection{History}

The \be approach for bulk data analysis was first deployed to  find confidential information on a set of 150
hard drives purchased on the secondary
market\cite{garfinkel:remembrance}. The program was refined and made
multi-threaded to keep up with the increased number of hard drives and
other storage devices collected during the construction of the Real Data
Corpus\cite{garfinkel:corpora}. A  study revealed
specific requirements that would be of use to  law enforcement; these
requirements were implemented\cite{GARFINKEL201356}.

During development, some BE users
indicated that they were excited by the project but wanted to have
their own, private label, proprietary version of the program that
incorporated specific non-public capabilities. Maintaining such
capabilities is  a complex undertaking. Instead, the decision
was made to refactor the program's code base, dividing the
functionality into three parts. The core architecture of applying
content-recognizing scanners to blocks of data and providing for
recursive re-analysis of that data was incorporated into the somewhat
improperly named \emph{bulk\_extractor API}. This module included
support for the program's configuration, passing data to scanners,
maintaining a set of scanners (the ``scanner set''), and a ``feature
recorder'' system for persisting the data found by scanners to the
file system for further analysis. Because this module was introduced
in version 1.3, this module was called \texttt{be13\_api}. We have since renamed it to \texttt{be2\_api} because of the significant incompatible changes associated with the 2.0 release.

The second part of \be was the program's main loop and all code
necessary for reading disk images. This part of the program reads data
in overlapping blocks and feeds the data to the API. When a user
requested the ability to have BE scan files in a file system (or
contained within a cloud-based storage system), this new capability
was readily added, with only a few lines of modification required to
the main program.

The third part of BE was the scanners themselves. Each scanner used
the same API and only the API: no scanner was privileged over any
other.  Written in C++,
scanners can be compiled and linked with the main program and the API. Alternatively, scanners can be embedded in shared
libraries (\texttt{.so} files on Linux and macOS, \texttt{.DLL} files on Windows)
and loaded at run-time. The scanners register not just their ability
to scan data and write to feature files, but their metadata,
configuration variables, and help messages: all are available using
standard command-line arguments. This allowed users to create and
deploy their own proprietary scanners.

BE1.6 shipped with 37 individual scanners; BE2 has 35 (\texttt{hashdb}
and \texttt{sceadan} both having been removed). The amount of effort to take an existing
digital forensics C or C++ library that extracts features from a block of data  and turn it into a BE scanner is typically less than an hour, provided that the library is already thread-safe.

\subsection{CLI and GUI}

BE was designed to be used with a command line interface (CLI) that performs batch
analysis on pretty much any kind of data that a forensic investigator
might have. The command-line user interface is straightforward: one
provides  the input file and an output directory, and BE
runs. (The program has over a hundred command-line
arguments to enable or disable scanners and to set configuration
variables, but all can be safely ignored in most cases.)

BE also has a graphical user interface (GUI) written in Java. Called ``BEViewer,'' the program's main feature is
viewing the ``feature files'' and carved data that BE produces. BEViewer can also run the BE program.

\subsection{BE in research}

The original BE feature extraction code was developed to search for
credit card numbers and other sensitive information on hard drives
purchased on the secondary market as part of a research study\cite{garfinkel:remembrance}, and the
program retained its use as an apparatus for digital forensics
research and experimentation for the following decade. For example,
the \texttt{sceadan} scanner was implemented for Beebe and Maddox's
Sceadan tool\cite{6567922}, and a \texttt{hashdb} scanner was
implemented for  Allen's  \texttt{hashdb}\cite{hashdb}. Incorporating
the experimental functionality into BE allowed testing  the Sceadan and
HashDB algorithms at scale, with large amounts of data, and to have the results
recording using the BE feature reporting system.  This may have made it
faster to develop these systems.

Building Sceadan and HashDB into the BE mainstream code had
negative impacts as well: doing so complicated building the program. It also added to
the cost of supporting BE, since users wanted to know if Sceadan or
HashDB were required for proper operation (they were not). Finally, these experimental
tools were not needed by the majority of BE users. Indeed, despite
multiple publications on fragment classification and hash-based carving (\cite{young:distinct,garfinkel:hashdb}), there is no evidence that this technology
was ever deployed into an operational environment.

As a result, we have removed support for these experimental systems
from BE2.0. Users who need this functionality can use BE's ability to
load plugins at runtime using shared libraries. Indeed, it is unclear
why the original developers of these scanners did not  implement them
as shared libraries.

\subsection{BE in education}

BE has  been widely used in digital
forensics education, as evidenced by the more than 400 videos on
YouTube that result from a search for the term ``bulk\_extractor.'' Many of these
videos  showcase the result of student projects using the tool.

We believe that BE is a successful tool in education because is easy
to use, runs on Windows, Mac and Linux platforms, and finds a wide
variety of forensic artifacts. For advanced students,
BE can easily be used as inputs into a wide range of student projects.

\subsection{BE in operational use}

Since its creation, \be has been used by  government agencies worldwide and private companies. In 2011 the program won a US Department of Defense Value
Engineering Award~\cite{bulk_extractor-ve}.

Because \be is distributed as open source software and because most investigations
are subject to confidentiality constraints, it
is difficult to establish how broadly the program is used. For
example, the program is included  on several digital forensics software
distributions, but it is not possible to know if people who run the
distributions from a bootable DVD or USB memory stick actually run the
\be program. We do know that the program is included as part of the Blacklight
digital forensics tool\cite{blackbag-agreements}, and  has also been used
incorporated into the BitCurator\cite{bitcurator} tool used by curators in the digital
humanities.

Stroz Friedberg's DFIR consulting practice has made use of \be in some
investigations. In a large incident response case, Linux servers with
XFS filesystems had been attacked, and no popular forensic tools could cope
with filesystem analysis of XFS. \be was used to triage these servers for
relevant indicators of compromise, allowing for rapid progress in the early
days of the investigation. In a well-known intellectual property theft case,
\emph{Waymo vs Uber}, \be was used as one of several processes to scour
forensic evidence of the defendant's engineering laptops for file names
provided by the plaintiff.

In summary, \be is a powerful tool that has been used for more than a
decade, with compelling anecdotes of usage, but we do not know how widely or regularly it has been used.

\section{Updating BE}\label{improvement}
\be is a legacy C++ program. The technique of using GNU~flex to
develop large regular expression ensembles dates to an unpublished 1989 MIT Media
Lab research project. The histogram engine dates to Garfinkel and
Shelat's 2003 project~\cite{garfinkel:remembrance}. The
producer-consumer threadpool was developed in 2008, and the underlying
scanner-based architecture with recursive reanalysis was in place by
2009. All of this was done with versions of C++ based on the
circa-1998 Standard Template Library (STL), well before the
ratification of the C++11 standard.

Part of the \be requirements study\cite{GARFINKEL201356} revealed that the application
needed to run in a wide variety of environments, including Microsoft
Windows, macOS, and several varieties of Linux. The program achieves
the necessary portability through the use of GNU autoconf\cite{Vaughan:1093139}, the POSIX API, and
the mingw\cite{mingw} compiler suite to produce the Windows executable. The
ability of these tools to provide portability to future operating
systems is less well developed, however, necessitating minor changes to the
configuration system or the \be source code to accommodate operating
system changes such as deprecated APIs and renamed \texttt{\#include}
files.\footnote{Sometimes compiling on a new operating system is more difficult than merely
changing filenames. A recent example: after a colleague upgraded a
build system, it was discovered that The
Sleuthkit compiled with the MinGW cross-compiler would no longer
function properly. It was discovered that between Fedora 31 and Fedora
34 the compiler's developers had
swapped MinGW's handling of the \texttt{printf} format specifiers \texttt{\%s} and \texttt{\%S}, from Windows semantics\cite{raymond}
to POSIX semantics\cite{broken}.
We avoided this problem in BE2 by depreciating the use of
\texttt{printf}, relying instead on the C++17 formatting primitives
which are consistent across platforms.}

Development of new \be features largely stopped in 2014. The one exception was an unfortunate fork of the \be
codebase when a developer added support for ``record carving.''\cite{record-carving} After some
analysis, the new record-carving scanners were incorporated into the
main \be codebase.

Nevertheless, software maintenance remained an ongoing concern: with
each new release of an open source operating system, the autoconf
system typically required some changes so that \be would compile on
the new system. By 2018, such changes were coming with alarming frequency.

We were also caught off-guard by the deprecation of Python version 2. The BE codebase always worked with both Python~2 and Python~3. However, when Python~2 reached end-of-life on January 1, 2020, at least one Python indicated that the stated compatibility with Python~2 meant that the program included Python~2 support, and this was deemed unacceptable. We resolved the issue by publishing a new version of \be that only advertised compatibility with Python~3.

\subsection{Upgrade Goals}
Based on this experience, in 2018 the decision was made
to embark on an orderly upgrade of \be to create version 2.0 This
section describes the upgrade goals.

\paragraph{Make the program easier to compile and maintain by relying on the
C++ standard} The primary reason for the \be upgrade was that the
program would no longer compile on modern open source operating systems. In part, this
was because \be predated the C++11 standard. Although the autoconf system is resilient, it has its limits,
and after six years of abandonment, they were beginning to show.

We were especially eager to rely on the C++ standard to provide platform-independence, because the
C++ standard is designed so that conforming code can compile in the
future on platforms that do not exist today. It does this by
specifying the \emph{version} of the standard to use when compiling
and linking the executable: C++11, C++14, C++17 and so-on.

Upgrading the existing code to a modern C++ standard required that we:

\begin{compactitem}

\item Choose a specific C++ standard.

\item Generally use C++ functionality rather than POSIX or Windows
  functionality.

\item Remove as many \texttt{\#ifdef} preprocessor directives as possible.

\item Replace code that the original authors had painstakingly written, debugged and
maintained with new code that used the C++ standard. This meant
that we might be introducing bugs into working code, so we needed to
have a better strategy for testing than the original authors.

\end{compactitem}

We first chose the C++14 standard, as a complete C++17 implementations were not widely
available when the upgrade started. We eventually migrated to C++17 for the
\texttt{std::filesystem} support, and because the project had dragged on for
so long that C++17 was widely available.

In 2021 we considered moving to C++20, but attempts to use specific
features met with failure, so BE2.0 uses C++17.

\paragraph{Improve reliability} The single most important aspect of a
digital forensics tool is that it must not crash before useful output.
Ideally a digital forensics tool will not crash at
all. However, if the tool does crash, there needs to be some way to
get useful output first, or to restart the tool to complete the
analysis.

Garfinkel's original BE requirements study identified the need to
never crash\cite{GARFINKEL201356}. Of course, the tool did crash from
time to time, so BE incorporated a system for restarting: when the
command-line tool crashed, if the program was re-run with exactly the
same command line arguments (by hitting up-arrow and return, for
example) the tool would carry on from where it had left off, skipping
the data that it was analyzing during the crash. For BE2, a goal was
to improve the program's reliability so that it really never did crash.

\paragraph{Simplify the code base} Although the
internal structure of \be was sound, it was needlessly complicated in
places, a result of 10 years' of evolution in the calling conventions
between the API and the individual scanners. For example, scanners in
the \texttt{be13\_api} took two arguments: a pointer to a structure called the
\emph{scanner parameters}, and a pointer to a second structure called
the \emph{recursion control block}. In \texttt{be2\_api} the relevant parameters from the recursion control
block were added into the scanner parameters structure, and that structure
is passed to the scanner as a reference to a C++ object, rather than a
pointer, because it is mandatory.
Unused options were removed, and options that were always
used together were combined.

\paragraph{Remove experimental code from the code base} \be
was initially developed to support digital forensics research, and
there was a significant amount of experimental, research code
contained therein. This code was removed for BE2: experiments
can continue, but they will be confined to using the plug-in system.

\paragraph{Make BE run faster}

Our final objective was to decrease the amount of time that the
program required to run. Initially,  we
wanted BE2 to run faster than BE1 on the same hardware. After
further analysis, we determined that BE2 should also take better
advantage of multiple processor cores without corresponding  need
for high-performance I/O systems: over the past ten years CPUs had
enjoyed much more speedup than disk subsystems.

To increase parallelism, we redesigned the multithreading system so that recursive processing could happen
in another thread. This allows utilizing more cores without requiring
improvements in the underlying I/O system. We also added
reference-counting garbage
collection to the memory management system so that a single buffer
could be processed simultaneously by multiple scanners, each in their
own thread: the memory is automatically freed when it is no longer
needed.

BE has the ability to process a directory of files. In BE1 each file was handled in its own thread. In BE2 all of the files are scanned in advance and then processed in order, each file being split into multiple pages, each page being processed in parallel with multiple threads. As a result, there are more opportunities for parallelism. (We also now recursively enumerate the  directories and files  within the specified directory using the portable C++17 methods, which further simplified the BE codebase.)

Another improvement in computing since the release of BE1 was the
widespread availability of serverless computing systems such as Amazon
Lambda and Microsoft Functions. Because BE effectively processes each
16MiB page independently, 1TB of evidence can be
processed simultaneously on 59,605 different VMs. This would allow
processing the 1TB disk image in 5~min or less \textit{without sampling}, a longstanding goal of this
project.\footnote{The goal of processing a 1TB disk in five minutes was
first proposed by a DARPA program manager in 2005.} Realizing this goal would require storing the 1TB
drive on a parallelized storage system that could accommodate tens of
thousands of simultaneous and independent readers. It would also
require the ability to scale Amazon Lambda from 0 to 59,605
simultaneous function executions instantaneously. In practice, Amazon
has configured AWS to scale gradually. So while adopting BE2 so that it could run under Amazon Lambda would
allow processing a 1TB drive in 5 minutes, it would not be the first
drive that received such a speedup: it would likely be the 10th or
20th. This experiment will not be realized until BE users have the
need for such capability, and are willing to pay to develop it.

Finally, we wanted faster compiles for developers, which meant simplifying the GNU \texttt{autoconf} script,
as this script runs single-threaded.

\subsection{Improving the Code Quality}

As part of refactoring the code base, we dramatically improved the
code quality of the underlying C++ code.

We started by reading most of  Stroustrup's
textbook\cite{c++}. More than a thousand pages long, we believe that few people read
this book in its entirety. Moreover, the book only covers through
C++11, and we were using C++14 (and then C++17). However, \be was based on a version of C++ that predates the C++11 standard, and the changes between that version and C++11 are dramatic compared to those that follow. The intimate familiarity that comes with reading such a text
allows one to make better use of the language's features that are at
the same time both more efficient and safer.

Next, we worked to improve the efficiency, safety and speed of BE's
fundamental memory management C++ class, the \sbuf. This class
represents a sequence of bytes that are read from evidence or decoded
from another \sbuf. The \sbuf tracks how the contained memory was
allocated (and thus, how it needs to be freed); provides accessor
methods that are type-safe, memory-safe and thread-safe; has
capabilities for making new \sbufs from disk files, slices of other
\sbufs, from new memory that is passed to a codec or decompressor; and
provides rich debugging capabilities.

To improve encapsulation and provide for better code re-use, we moved many functions from scanners and the BE framework into the
\sbuf class. For example, the \sbuf can now compute its cryptographic
hash in a threadsafe manner and cache the results so that they can be
used elsewhere in the program: hashes are computed as needed, a
form of lazy evaluation. The \sbuf class now implements string-search
and several other performance-critical functions. Implementing them
within the \sbuf class allows the arguments to be validated for memory
safety \emph{once}, and then unsafe code to be used within the \sbuf
implementation. As a result of moving functionality into the \sbuf, we
were able to eliminate virtually all raw memory references throughout
the rest of BE.

Other improvements in the code base include:

\begin{compactitem}
\item We moved common code out of the scanners and into the \texttt{be2\_api} framework. For
example, rather than each scanner having options for setting its carve
mode, the framework understands how to set them for any named
scanner.

\item We simplified the API, combining functions and
  methods with nearly identical functionality. For example, \sbuf
  previously had two functions that could map a file into an \sbuf: one that
took filename already open and the open file descriptor, another that just took
the filename and opened it. We eliminated the API call that took an open file descriptor,
and modified the source-code so that the file was opened by the \sbuf
method, and never by the caller.

\item We added explicit phases into the API where the scanners can
  allocate and free memory. Now scanners are expected to deallocate
  all memory that they allocate during the run, rather than allowing
  the operating system to discard the memory when the process
  exits. This allowed
  us to find memory leaks in our unit tests that otherwise would have been missed.

\item Whereas previously many strings were passed by reference as \texttt{const std::string \&},
there are now passed by value as a \texttt{std::string}. This necessitates a
string copy, but it is not a meaningful impact on performance,
especially when compared with the improved safety against possibly
using an invalidated reference. This decision simplified code and
resulted in the elimination of several use-after-free errors.

\item We enabled \sbuf child tracking, meaning that each \sbuf counts
  how many child \sbufs it has. (A child \sbuf is one that shares a
  portion of the parent \sbuf's memory.) Previously this was turned off
because of a bug in which not all children were properly registered
when they were created and de-registered when they were deleted. Once we defined the
clear policy described above, we were able to find the bug! Now we assume that all allocated \sbufs are freed and throw an exception if that is not the case, which allows memory leaks to be rapidly identified during software development.

\item With the above tracking of \sbufs, we now separately count the total number of \sbufs allocated and freed and validate that there are zero \sbufs remaining when the main analysis is complete. Once again, this allows us to rapidly identify memory leaks during development.

\item We now define a clear allocation/deallocation policy for all
objects in memory. Special attention
has been made to implementation of C++ move operators, allowing the
compiler to realize increased efficiency by using use them instead of a copy and delete operator.

\item Code that was \texttt{\#ifdef}'ed for Windows, macOS and Linux was
  replaced where possible with calls to the C++17 library. In
  particular we made extensive use of the \texttt{std::filesystem} class. The
  result of these changes made the code smaller and easier to validate.

\item We likewise replaced many pre-processor \texttt{\#define} statements with C++ inline static constants whenever possible. This makes the values available to the debugger, and makes the code easier to understand.

\item We eliminated global variables used to track state. The only use
  of global variables that remain are static tables that are used for
  the precomputed value of CPU-intensive functions. The code that uses these variables now checks to
verify that they have been initialized and throws an exception if they
have not. Essentially, they are now singletons. It would be nice to
change the memory protection of these variables to read-only, but that
cannot be done in a portable manner and would require that the
variables have their own memory pages.

\item In many cases we have removed return codes that must be checked to detect
errors. Instead, we use the C++ exception mechanism to signal and
catch error conditions.

\item BE1.6 used \emph{gcc} compiler intrinsics for atomic
increment in some locations, but made broad use of mutexes to protect
variables shared between threads. In BE2 we have eliminated many explicit mutexs and replaced them with the C++
\texttt{std::atomic<>} template.

\item We removed the legacy POSIX \texttt{getopt} processing and replaced it
  with \texttt{cxxopts}\cite{cxxopts}, a command line option processing module that is
  reentrant and does not make use of global variables. This was
  necessary to allow unit tests that tested the option processing.

\item We enabled all compiler warnings, not simply those enabled with \texttt{-Wall}.

\end{compactitem}

The results of these changes:

\begin{compactitem}
\item The BE2 configuration script now runs in 16 seconds on our
  reference Mac mini, instead of 25 seconds for the BE1.6 configuration
  script. This is not a significant improvement for users, but it is
  for BE developers. The compile time for both is 32 seconds using \texttt{make -j12} on the 6-core system.

 \item
  We were able to reduce the C++ code base by roughly ten thousand
  lines, or ~17\%, even accounting for the lines we added for the new unit
  tests. The program is now roughly 46~thousand lines of C++ and GNU
  flex code.

\end{compactitem}

\subsection{Dynamic Analysis with Unit Tests and Test Coverage}

As hinted  above, despite widespread use, \be lacked a modern
approach to testing. Specifically, the \be codebase was  devoid
of systematic unit tests. Instead, there were test programs
that were occasionally manually run from the command line for comparing the output with
results of a previous run: if more features were
extracted, the program was not obviously broken.

We implemented unit tests for all
levels of the \be source code. We reviewed the list of C++ unit test frameworks on
Wikipedia and chose  Catch2\cite{catch2}, which has support for test scaffolds, implements a minimal
CLI, can test
for the presence (or absence) of thrown exceptions, and appeared to be well supported and  maintained.

We also  enabled AddressSanitizer~\cite{10.5555/2342821.2342849} by default on
our development system. (We enabled
ThreadSanitizer\cite{10.1145/1791194.1791203} and found several thread sharing errors, but we also encountered a false
positive due to a conflict between one of its heuristics and our
multi-threading paradigm, preventing us from leaving it enabled by default.)

We started with unit tests for the \texttt{be2\_api} framework. We generally wrote unit
tests as the new interfaces were designed and implemented, combining
the creation of each new test with related refactoring. We decided to track and systematically increase the \emph{code
coverage} of the unit tests.
We used the popular
CodeCov.io website to display the code coverage of the unit tests.

Creating code coverage reports for C++ was straightforward: we re-ran
\texttt{./configure} specifying additional compiler flags and libraries, then run
a post-processing tool after the unit test runs, and finally run
CodeCov's script to upload the report to the website. After we got
this working, we then integrated it with GitHub's ``Actions,'' so that
the unit tests would automatically be run and coverage reports
uploaded after every commit to GitHub or pull request.

After all of the new and refactored code had unit tests, we examined
the code coverage reports to determine which pieces of legacy code
were not covered by the newly written unit tests. We established a target code coverage of 60\%. In some cases,
legacy code \emph{was} covered by the new tests, because the new code
called the old code. But for roughly two thirds of the code, there was
no coverage by unit tests. For this legacy code, at first writing unit tests
seemed largely like a compliance
exercise---after all, \be had been in use for more than a decade, so
we thought that all of the significant bugs, such a memory allocation
errors, off-by-one errors, and so on, were gone from the code base.
However, the act of writing the unit
tests forced us to clarify internal documentation, simplify
internal implementations, and in some cases we were able to eliminate
legacy code that was no longer being used. To paraphrase the immortal
Steve Jobs, the most reliable piece of code, the piece of code that
you never need to test, is the line of code that you don't write---or
in this case, the line of code that you remove from your legacy programs. In total, more than ten thousand lines of C++ code was removed between BE1.6 and BE2.

\subsection{Removing Functionality}

In addition to removing experimental functionality, we improved
performance of \be by disabling some functionality that would normally
never be executed. In some cases the functionality can be
re-enabled with command-line switches; in other cases it cannot.

Functionality that was disabled includes:
\begin{compactitem}
\item We added flags to the description of scanners so that specific
  scanners that look for in-memory artifacts or disk-based artifacts
  will never be called to process the results of the majority of
  scanners that initiate a recursive reanalysis. For example, it makes
  little sense to look for NTFS directory entries in a decompressed gzip
  stream---acknowledging that this means we will not scan for
  filesystems on gzip-compressed disk images.

  \item By default, we now disable the hiberfile scanner (xpress
    decryption), because we lacked test vectors that could be used
    to demonstrate the correctness of our implementation, and because
    Windows may no longer be using the compression algorithm that we
    have implemented.

\item We disabled (by default) scanning for 192-bit AES keys in
  memory, because in practice AES is rarely used in its 192-bit mode.
\item We disabled (by default) xpress decompression, as other
  algorithms are now used to compress swap memory.
  \end{compactitem}

All of the features that are disabled by default can be re-enabled
with a command-line switch.

We also removed key functionality from the tool that we had determined was not being used:

\begin{compactitem}

\item We used internet search engines to see if some of the program's more
  obscure command-line options are being referenced in open source
  programs, scripts, or even on blog entries that provided tutorials
  for using BE. Obscure options that were unused by the user community
  were eliminated.

\item To the best of our knowledge, no one (other than the original
developer) ever used BE's shared library to let the
program's scanner system be called from C++ or from Python, so we
dropped support for that. (It could be trivially added in the future.)

\item The stand-alone BE test program that only scans a single
file was dropped as additional code that did not need to be
maintained. Instead, we now have the unit tests.

\item The ability to load scanners as shared libraries at startup has
  not been updated for BE2, although this update is trivial and will
  be implemented if users request it.

\end{compactitem}

Finally, we moved the BE2 Java user interface out of the BE repo and
into its own that has the BE repo as a sub-module.  While
the Java GUI runs just fine with the BE2 engine, the  build system has
changed significantly. Moving the Java GUI into its own git repo
allows us to better isolate the two build systems.

\subsection{Incompatible Changes}
Despite our efforts to retain full compatibility between BE1 and BE2, we needed to introduce a few minor incompatible changes were required in the interest of correctness and modernization:

\begin{compactitem}
    \item BE feature files are utf-8, but some of the information in them is binary and must be escaped. In BE1 non-Unicode characters were present and escaped in octal. In BE2 non-Unicode characters are escaped in hexadecimal.
    \item A persistent problem is how utf-16 features should be represented in the utf-8 feature files. BE1 presented UTF-16 as octal-escaped values, which was hard to read. BE2 converts utf-16 into utf-8 in the second (``feature'') column, but leaves the features as (escaped) utf-16 in the third (``context'') column.
    \item BE2 properly reports the start of features that are within ZIP-decoded data blocks, (see \S\ref{validation} for a detailed discussion).
    \item BE1 computed the MD5 hash code of forensic media that it processed; BE2 uses SHA-1.\cite{sp800-131a-r2}

\end{compactitem}

\subsection{Performance Tuning}
Despite the effort to eliminate all memory copies, an interim  version
of BE2 was dramatically slower than BE1.6. For example, scanning the
2009-domexusers\cite{garfinkel:corpora} disk image on a 6-core Mac mini required approximately 10 minutes
with BE1.6, but took 70 minutes with the development version .

BE has long had the ability to measure each scanner's  contribution to runtime. Specifically, it keeps counters (in
\texttt{std::atomic<>} variables) of how many times each scanner is called and
how many nanoseconds it spends executing. These counters became more
accurate in BE2.0, with the decision to queue the recursive processing
of \sbufs longer than 4K to another thread. Looking at these counters
we saw that just three scanners (\texttt{rar}, \texttt{net}, and \texttt{aes}) were responsible
for the vast majority of the time spent scanning.

Each of these scanners has a hand-coded loop that scans through the
memory image looking for a magic number. The loop had been implemented
making a new \sbuf for each location.  Analysis of a 2GB disk image required creating
over 3 billion \sbufs!  The first improvement we made was to implement
the validator so that instead of validating the first position in the
\sbuf, it would take an offset. This eliminated the need to create a
new \sbuf for each offset. (Creating new \sbufs is cheap, but not free!) Once the magic number was found, a new \sbuf was created, so
as to take advantage of the algorithmic simplification. Additional
improvements were realized by moving the search for magic numbers into
the \sbuf implementation itself, so that it could be performed with
\emph{memchr}.

Once these changes were made, the \texttt{rar} scanner was no longer the slowest. Now the slowest scanners were \texttt{net}, \texttt{aes}, and the flex-based scanners \texttt{email} and \texttt{accts} (but not the other flex-based scanners, curiously enough). More than a month was spent going through these scanners line-by-line in an effort the determine the precise C++ statements responsible for the slow-down.
The end results is that BE2 is now substantially faster than BE1 in all cases (see \tabref{speedup}).

\begin{figure*}
{\small
\begin{Verbatim}
bulk_extractor 1.6:
456536-ZIP-0-MSXML-9 iwork09.comkeynote_comment@iwork09.com   keynote@iwork09.comkeynote_comment@iwork09.com
456536-ZIP-1255117  keynote@iwork09.com.                      nk href="mailto:keynote@iwork09.com?subject=">

bulk_extractor 2.0:
456596-ZIP-0-MSXML-9 iwork09.comkeynote_comment@iwork09.com   keynote@iwork09.comkeynote_comment@iwork09.com
456596-ZIP-1255117   keynote@iwork09.com                      nk href="mailto:keynote@iwork09.com?subject=">
\end{Verbatim}
}
\caption{Comparing the email.txt feature file output for BE1 and BE2 for the disk image nps-2010-emails. The left column indicates the forensic path, the middle column the feature, and the right column shows the feature in context. The first line shows an email address extracted from a Microsoft XML file that is within a ZIP file. The second shows an email addresses found within the ZIP file but not decoded by the Microsoft XML text decoder. Notice that the offset of the ZIP file is 60 bytes different; in the BE2.0 file this indicates the starting location of the ZIP header. The output has been reformatted for legibility.}\label{path-offsets}
\end{figure*}

\section{Validation}\label{validation}

``A program that has not been specified cannot be incorrect; it can only be surprising.''\cite{surprising}

We performed two kinds of validation on BE2: correctness and throughput. For correctness, we wanted to validate that BE2 produced results that were as good as the results of BE1. For throughput, we wanted BE2 to be at least as fast as BE1.

\subsection{Correctness}

When we found differences between the output of BE1 and BE2, some were cases in which BE2 was correct. In these cases, it appeared that the
BE1 output had never been validated in detail. Most of these had to do
with the location of recursively-analyzed features in the feature
file.

For example, the BE forensic path \texttt{456536-ZIP-1255117} is read to mean
that there is a feature that is located 1255117 bytes into a
inflated ZIP stream that is itself located 456536 bytes from the
beginning of the disk image. With the disk image nps-2010-emails\cite{garfinkel:corpora}, BE1
reported the ZIP stream beginning at 456536, but BE2 reports the same
ZIP stream beginning at location 456596. The 60-byte difference is the
result of the ZIP header. BE2 correctly reported that the ZIP stream
began at 456596 because the address was tracked automatically by the
revised memory allocation routines that tracked the location of the
sliced buffer that was handed to the decompressor. In BE1, the
address was computed with explicit code, and that explicit code
(rightly or wrongly) reported the location of the ZIP segment header,
rather than the ZLIB-deflated stream. (See figure~\ref{path-offsets})

We discovered this specific error writing a unit test to test the forensic path printer---the part of \be that reads a forensic path and provides a hexdump of the contents of the evidence so indicated. Although this code had been in use for more than 10 years in the \be GUI, apparently it had never worked properly for the ZIP scanner, and none of \be's users had ever reported it not working. (The GZIP scanner reported forensic paths correctly.)

Many of the code paths in the BE1 code base were painstakingly
developed on specific test cases, but those test cases were not added
to the code base as unit tests. For example, the \texttt{net}
packet scanner\cite{beverly:ipcarving}
could carve IPv4 and IPv6 packets as well as recognize in-memory TCP
header structures from Microsoft Windows memory dumps.
The part of the
scanners that accessed raw memory also received significant rewrites to go
through the new \sbuf API. We then wanted to validate that the
rewritten scanners had the same functionality as the old ones. The
only way to do this was by assembling specific test cases for each
data type---and adding them to the code base. This is something that
wasn't done originally.
Those test cases are now parts of the \be code base and the code is validated on every commit. This turned out to be invaluable for maintaining correctness during the performance tuning efforts described in the next section.

\subsection{Throughput}
It is straightforward to measure the speed with which \be processes a
disk image or other form of electronic evidence. Explaining variations
in speed is significantly harder. The time that \be spends processing evidence is highly dependent upon the contents. A disk that contains many compressed archives will take longer to process because each compressed run of bytes will be decompressed and recursively re-analyzed. A disk that is filled with JPEGs will analyze quickly, but if carving is enabled (the default), each JPEG will be copied off. (However, if carving mode is set to 2, only the JPEGs that had to be decompressed or otherwise decoded---the JPEGs typically missed by other carving tools---will be copied.)

BE also incorporates many techniques to discard data before applying the full recursive analysis. For example, duplicate data is typically not analyzed a second time. Likewise, pages that consist of a repeating n-gram (e.g. \texttt{ABCABCABC...}) will not be analyzed. Scanners contain flags in their metadata that determines if such analysis is desired.

Another factor in performance is the computer on which the program is run. The number of CPU cores, the amount of RAM, the speed of that RAM and the speed of the I/O system all impact throughput. And all of these factors interact with the evidence under examination: a disk image that has a lot of blank and repeated sectors will benefit more from a faster I/O system, while a disk image with a lot of complex data structures will benefit more from additional cores.

Therefore, throughput and benchmark results in general are best reported using evidence that is ecologically valid\cite{ecologically-valid}, such as an actual disk image. Although such media are commonly used in software development and in internal benchmarking, these media tend not to be publicly released due to privacy reasons.

\begin{table*}
  \begin{tabular}{ll|r||r|r||r|r||l}
                  &                       & \multicolumn{1}{r}{Scanners:}      & \multicolumn{2}{c||}{29} & \multicolumn{2}{c||}{30 + AES192} \\
\cline{4-5}\cline{6-7}
\textbf{Computer} & Disk Image (+ config) & BE1.6 & BE2 & Throughput  & BE2 & Throughput \\
\hline\hline
\multicolumn{8}{l}{\textbf{MacBook Pro (Retina, 13-inch, Late 2013) 2.8 GHz Dual-Core Intel Core i7; }}\\
\multicolumn{8}{l}{\hspace{1pc}\textbf{16 GiB  1600 MHz DDR3; 2 physical cores (4 with hyperthreading); macOS 11.6.3}}\\
& nps-2009-ubnist1 & \SI{140}{s} & \SI{109}{s} & 128\% & \SI{120}{s} & 117\%\\
& nps-2009-domexusers          & \SI{1420}{s} & \SI{837}{s} & 170\% & \SI{1208}{s} & 118\%\\
\hline
\multicolumn{8}{l}{\textbf{Mac mini (2018) 3GHz 6-core i5; 2667 MHz DDR4; macOS 12.1}}\\
& nps-2009-ubnist1    &  \SI{43}{s}  &  \SI{35}{s} & 123\%  & \SI{33}{s} & 130\% \\
& nps-2009-domexusers & \SI{428}{s}  & \SI{319}{s} & 134\%  & \SI{428}{s} & 100\%\\
\hline
\multicolumn{8}{l}{\textbf{MacBook Pro (16-inch, 2021) Apple M1 Pro 10 core;  32GiB RAM; macOS 12.1}}\\
& nps-2009-ubnist1 &  \SI{20}{s} &  \SI{16}{s}  & 125\% & \SI{17}{s} & 118\%\\
& nps-2009-domexusers & \SI{221}{s} & \SI{126}{s}  & 175\% & \SI{172}{s} & 128\%\\
& nps-2013-2tb & \SI{20142}{s}  & \SI{10944}{s} & 184\% & \SI{11184}{s} & 180\%\\
\hline\hline

  \end{tabular}
\caption{Clock time comparison of running  BE1.6 and BE2 on a variety
  of hardware and software configurations. All executables are compiled -O3. Times for the
  nps-2009-ubnist1 and nps-2009-domexusers are average of three
  runs. The nps-2009-ubnist1 and nps-2009-domexusers read and write to
the system SSD, while the nps-2013-2tb reads from the system SSD and
writes to an external USB3 HD due to storage considerations. BE1.6
speeds reported for runs
with the standard 30 default scanners enabled: accts, aes, base64,
elf, email, evtx, exif, find, gps, gzip, hiberfile, httplogs, json,
kml, msxml, net, ntfsindx, ntfslogfile, ntfsmft, ntfsusn, pdf, rar,
sqlite, utmp, vcard, windirs, winlnk, winpe, winprefetch and zip. BE2
for runs with the standard 29 scanners enabled (hiberfile is
disabled) and with AES192 key searching disabled, and with the BE1.6
configuration that adds hiberfile and AES192 key searching. The Apple
M1 Pro 10 core processor has 8 ``performance'' cores and 2
``efficiency'' cores. ``Throughput'' is normalized to the speed of
BE1.6 on the same hardware with the same disk image; a throughput of
200\% means that the disk image will be analyzed in half the time.}\label{speedup}
\end{table*}

In Table~\ref{speedup} we report the performance of BE1.6 and BE2.0 with three reference disk images from the DigitalCorpora collection, running on three different reference computers. The disk images are \textit{nps-2009-ubnist1}, a 2.1GB disk image of a bootable USB drive running Ubuntu Linux; \textit{nps-2009-domexusers}, a 42GB disk image of a Microsoft Windows system that was used by several individuals in a lab, and \textit{nps-2011-2tb}, a 2.0TB disk image containing the entire GovDocs1 corpus and several other of the DigitalCorpora reference disk images. All were made at the Naval Postgraduate School between 2009 and 2011 and are hosted on the DigitalCorpora website.

We report performance using three Apple Macintosh computers. Both
BE1.6 and BE2.0 were compiled on the computer on which the benchmark
was run with the current \texttt{llvm} compiler provided by Apple. All
compilation was done with \texttt{-O3}, with both AddressSanitizer and
ThreadSanitizer disabled. We report BE1.6 and BE2 with the default
analysis. In this configuration 30 scanners are enabled for BE1.6 but
BE2 disables hiberfile and AES192 key searching. For this reason, we
also report BE2 with the BE1.6 configuration.
As can be seen, BE2 is faster than BE1.6 in nearly every case,
although the speedup is more pronounced on the faster, more modern
hardware with more cores.


\section{Recommendations and Future Work}\label{recommendations}

This multi-year exercise shows the value of updating tools that appear to be working and bug-free to use current software engineering practices. We recommend a scrub of all modern digital forensics tools, as rewriting these tools will likely make them faster and more reliable.

Reading Stroustrup's book was time consuming preparation for this
project, but well worth the investment. We experienced a similar
benefit from reading the entire Python reference manual prior to
embarking on a large-scale Python project. We recommend detailed reading of all developer documentation for implementation languages and tools.
Organizations investing in digital forensics research and tools
should also be prepared to invest for the long-term, to provide for
maintenance, adaptation, and growth of promising tools, as well as focused
attention for developers.

We were stunned by the improvement in code quality that came from the pursuit of 60\% unit test code coverage. We were also surprised by the power of AddressSanitizer in finding a wide variety of bugs. We recommend adopting test-driven development\cite{beck:tdd} and test-driven refactoring\cite{fowler:refactoring} as a primary tool, and always enabling AddressSanitizer during the development process.

The dramatic speed of C++ compared to Python is a clear incentive to use this language for speed-critical applications. However, given the lack of C++ programmers in the digital forensics community, it is clear that BE requires an interface to allow Python scanners to be called. Because Python is not thread-safe, a separate Python interpreter will be required for each analysis thread. We recommend using C++ with well-designed classes to provide memory safety, and providing Python-based APIs to access their functionality.

We achieved a 61\% code coverage for the \texttt{be2\_api} but only 47\% for the BE2 code base (excluding the API). Clearly there is still room for improvement here.

Finally, the increased use of filesystem-level compression and encryption, combined with the use of the TRIM command on SSDs, means that the bulk data analysis of raw storage devices is likely to yield less data in the future than a systematic extraction of bulk data from resident files. That is, running BE2 with the -r (recursive) option on a mounted file system may one day yield more useful information than running it on the raw device. Ideally it would be possible to run BE2, keep track of the sectors that were scanned, and then process the remaining sectors raw. Another approach would be to perform two passes: one of the mounted files, and another of the raw device. Evaluation of these strategies is left as future work.

\def\urlprefix{}
\bibliography{references,garfinkel}

\end{document}